\documentstyle[prl,aps]{revtex}

\input{epsf}

\newcommand{\pmin}{$P_{\mbox{\footnotesize{min }}}$}

\begin{document}

\twocolumn[\hsize\textwidth\columnwidth\hsize\csname @twocolumnfalse\endcsname  
\title{Coherent transport in a two-electron quantum dot molecule}
\author{C.~E.~Creffield and G.~Platero}

\address{
Instituto de Ciencia de Materiales (CSIC), Cantoblanco, E-28049, Madrid,
Spain}

\date{\today}

\maketitle

\begin{abstract}
We investigate the dynamics of two interacting electrons confined
to a pair of coupled quantum dots driven by an
external AC field. By numerically integrating the two-electron
Schr\"odinger equation in time, we find that for certain values of the
strength and frequency of the AC field the electrons can become
localised within the same dot, in spite of the Coulomb repulsion between them.
Reducing the system to an effective two-site model of Hubbard type
and applying Floquet theory leads to a detailed understanding of this
effect. This demonstrates the possibility of using appropriate
AC fields to manipulate entangled states in mesoscopic devices on extremely
short timescales, which is an essential component of practical schemes for
quantum information processing.
\end{abstract}
\vskip2pc]
\narrowtext
\bigskip 

A set of electrons held in a semiconductor quantum dot is conceptually
similar to a set of atomic electrons bound to a nucleus, and for
this reason these structures are sometimes termed ``artificial atoms''
\cite{kastner}.
To extend the atomic analogy further, we can consider joining
quantum dots together to form ``artificial molecules''. The simplest
example consists of two coupled dots containing a single electron.
The dynamical properties of this system when driven by an AC field
have been extensively studied \cite{shirley}, and the technique of
Floquet analysis \cite{floq} has proved to be particularly effective.
The pioneering work in Ref.\cite{hanggi_prl} first noted the dramatic
effects on the tunneling rate arising from crossings and anti-crossings
of the Floquet quasi-energies, and the term {\em dynamical localisation} was
coined to describe the phenomenon in which the AC field appears to trap
the electron within just one dot over long time scales. Subsequent analytic
and numerical work \cite{holthaus}, \cite{hanggi_epl} showed that for high
frequencies this localisation occurs when the ratio of the field strength
to the frequency is a root of the Bessel function $J_0$.

Adding a second electron to the coupled dot system introduces
considerable complications, as for realistic
devices the Coulomb interaction between the electrons cannot be
neglected, and a single-particle analysis is not applicable.
Achieving an understanding of the dynamics of this system is, however,
extremely desirable, as the ability to rapidly control the localisation
of electrons using AC fields immediately suggests
possible applications to quantum metrology and quantum information processing.
In particular, manipulating a pair of entangled electrons on short
timescales is of great importance in the rapidly developing field of quantum
computation. Although numerical investigations have been made of the rich
behaviour displayed by interacting electron systems driven by an
AC field \cite{tamb_epl}, there is, at present, little
understanding of the observed effects beyond phenomenology and
numerical experimentation.

We address this problem here by applying the
Floquet formalism to a system of {\em interacting} particles.
We initially consider a realistic model of two interacting electrons
confined to a pair of coherently coupled quantum dots, and study
its response to AC fields by a numerical method in which the Coulomb
interaction is treated exactly. We find the surprising result that even in
the presence of strong inter-electron repulsion, suitable AC fields
can localise both electrons within the same dot. We further show
that this effect is also exhibited by a simple two-site model,
obtained as a reduced form of the quantum dot system.
As in the case of the single-particle model, we find
Floquet theory to be an extremely powerful tool,
and we succeed in generalising the single-particle
solution \cite{shirley},\cite{holthaus} to include interactions,
from which we can easily and accurately
identify the parameters of the AC field for which localisation occurs.

\begin{figure}[tbh]
\centerline{\epsfxsize=80mm \epsfbox{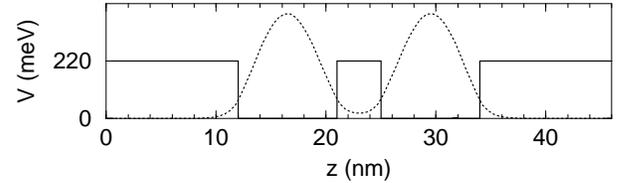}}
\caption{Geometry of the double-dot system.
The dotted line plots the charge density of the ground state wavefunction.}
\label{potential}
\end{figure}

We study a quasi-1D structure, in which the linear dimensions of the
quantum dots in the $(x,y)$ plane are much smaller than their length.
The energies associated with transverse excitations are thus much
higher than for longitudinal motion, and can be neglected. The dots are
separated by a potential barrier which regulates the degree of tunneling
between the dots. Strongly coupling the dots allows the electrons to form
extended states (``molecular orbitals'') over the entire double-dot system,
which can be explicitly seen in transport experiments 
\cite{blick},\cite{ooster}. Within the effective mass
approximation, the Hamiltonian for the system is given by:
\begin{eqnarray}
H = &-& \frac{\hbar^2}{2 m^{\ast}}
\left( \frac{\partial^2}{\partial z_1^2} + \frac{\partial^2}{\partial z_2^2}
\right) + V(z_1) + V(z_2) \nonumber \\
&+& \frac{e^2}{4 \pi \epsilon} V_C(| z_1 - z_2 |) \
-e E(t) (z_1 + z_2)
\label{hamiltonian}
\end{eqnarray} 
where $z_1, \ z_2$ are the spatial coordinates of the two electrons,
$m^{\ast}$ is the effective mass and $\epsilon$ is the effective
permitivity. GaAs material parameters are used,
with $m^{\ast} = 0.067 m_e$ and $\epsilon = 10.9 \epsilon_0$.
$E(t)$ is the external AC field, $E(t) = E \cos(\omega t)$.
To simplify the numerical treatment, the system was placed within
an infinite square well of length $L$ to prevent the electronic
wave-function from leaking out of the dot structure. So that
these infinite barriers did not unduly influence the properties of the
quantum dots, thick buffer zones were included between these
barriers and the walls of the quantum dots so that the electronic
wavefunction effectively decays to zero before reaching the
barriers. The confining potential, $V(z)$, is plotted
in Fig.\ref{potential}. The Coulomb potential used, $V_C(r)$, had
the form:
\begin{equation}
V_C(r) = \frac{1}{\sqrt{r^2 + \lambda^2}},
\label{coulomb}
\end{equation}
where $\lambda$ is a measure of the transverse width
of the structure (taken to be 1 nm in this investigation).
This form for $V_C$ reproduces the normal $1/r$ fall-off
of the Coulomb potential, but is non-singular in the limit $r \rightarrow 0$.

To study the time evolution of the system we used the eigenstates
of the static Hamiltonian, $H_0$, as a basis, since expanding in
these states is a well-controlled procedure, and also ensures that
correlation effects arising from the Coulomb interaction are automatically
encoded within each basis function. We note that as Eq.\ref{hamiltonian}
contains no spin-flip terms, there is no mixing between the singlet and
triplet sub-spaces. In particular, if the initial state has a definite
parity this symmetry is retained throughout its time
evolution, and consequently only basis functions of the same symmetry
need to be included in the expansion. To obtain the
eigensystem of either the singlet or triplet sub-spaces, we employed
the Lanczos technique described in Ref.\cite{charles}, which is particularly
efficient for systems in which the interaction is diagonal in real-space.

The initial state used was the ground state
of $H_0$ (a singlet). In the absence of the external electric field
this would have a trivial time evolution, simply acquiring a phase.
Applying the field, however, causes the initial state to evolve into a
superposition of eigenstates:
$| \psi(t) \rangle = \sum c_n (t) \ |E_n \rangle$.
Substituting this expansion into the time-dependent Schr\"odinger equation
yields a first-order differential equation for the expansion
coefficients $c_n$:
\begin{equation}
i \hbar \frac{d c_n}{d t}  = c_n(t) E_n - e E(t) \sum_{m=1}^{M}
F_{n m} c_m(t),
\label{derivs}
\end{equation}
where $E_n$ is the n-th eigenvalue of $H_0$, and $F_{m n}$ are the
overlap integrals of the dipole operator,
$F_{m n} = \langle E_m | (z_1 + z_2 ) | E_n \rangle$ .
A fourth order Runge-Kutta method was used to time-evolve the
expansion coefficients using Eq.\ref{derivs}. The number of basis states
used in the expansion, $M$, required for convergence
depended on the strength of the electric field, and values of $M=50$ were
required for the strongest field considered.

A similar model was studied recently in Ref.\cite{tamb_epl} for a
larger system size of $L = 100$ nm. An important advantage of
considering a smaller system, however, is that
finer structure can be resolved in the results due to
the larger spacing between the energy levels.
Following Ref.\cite{tamb_epl} we define a conditional probability
function, which gives an extremely useful description of the state of the
system. The probability that one particle is in the right dot while the
other is in the left is given by $P^{RL}(t)$:
\begin{equation}
P^{RL}(t) = 2 \int_R dz_1 \int_L dz_2 | \psi(z_1, z_2) |^2,
\end{equation}
where the notation $R$ / $L$ signifies that the integration is taken
over the right/left quantum dot. $P^{RL}$ takes a value of one for
maximally delocalised states (when one electron is in the right dot
and the other is in the left), and is zero
for localised states when both electrons occupy the same dot.
The initial state plotted in Fig.\ref{potential}
is highly delocalised, and has the value $P^{RL} = 0.849$.
Our investigation consists of evolving this state through a time
period of 18 ps while measuring $P^{RL}(t)$.
We term the minimum value of $P^{RL}$ attained during this period \pmin,
and use this to quantify the degree to which the AC field brings about
localisation.

\begin{figure}[tbh]
\centerline{\epsfxsize=80mm \epsfbox{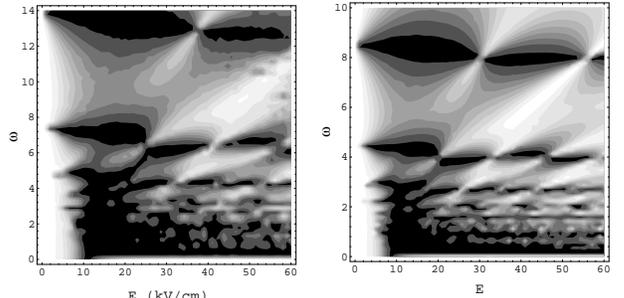}}
\caption{\pmin as a function of the strength $E$ and
energy $\hbar \omega$ of the AC field:
(a) for the interacting quantum dot system ($\hbar \omega$ in units of meV)
(b) for the Hubbard model with $U=8$ (both axes in units of $\tilde t$). }
\label{interact}
\end{figure}

\begin{figure}[tbh]
\centerline{\epsfxsize=80mm \epsfbox{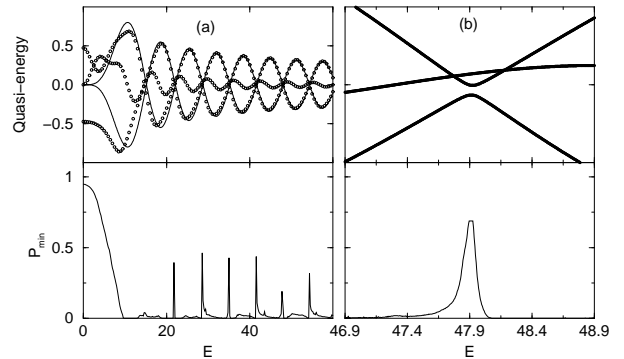}}
\caption{(a) Quasi-energy spectrum for the two-site model for
$U = 8$ and $\omega = 2$, circles = exact results,
lines = perturbation theory,
(b) magnified view of exact results for a single anti-crossing.
Beneath are the corresponding plots of \pmin.}
\label{quasi}
\end{figure}

In Fig.\ref{interact}a we present a contour plot of \pmin
as a function of the parameters of the AC field. Dark areas correspond
to low values of \pmin, indicating a high probability that
both electrons are occupying the same dot. Surprisingly this
can occur even at weak field strengths,
despite the presence of the Coulomb interaction.
We note that the dark areas form horizontal bands, indicating
that for various ``resonant'' values of $\omega$, localisation can
be produced over a wide range of fields. The spacing of the bands
decreases with $\omega$, and for values of $\hbar \omega < 2.8$ meV
the structure is too fine to be resolved. Between these
bands strong localisation is not produced. These results are qualitatively
similar to those of Ref.\cite{tamb_epl} but show finer detail, allowing
us to additionally observe that these bands are punctuated by
narrow zones in which the field does not create localisation.
Their form can be seen more clearly in the cross-section of \pmin
given in Fig.\ref{quasi}a, which reveals them to be narrow peaks.
These peaks are approximately equally spaced along each resonance,
the spacing increasing with $\omega$.

We emphasise that these results are radically different to those
obtained for non-interacting particles. In this case an analogous
plot of delocalisation shows a fan-like structure
\cite{hanggi_epl}, \cite{wang}, in which localisation occurs along lines
given by $\omega = E / x_j$, where $x_j$ is the $j$-th root
of the Bessel function $J_0(x)$. As a test of our method, we repeated
our investigation for a quantum dot system in which the inter-electron
Coulomb repulsion was set to zero, and found that the fan structure was
indeed reproduced.

To account for these results it is thus necessary to go
beyond the non-interacting case. We consider a highly simplified model
in which each quantum dot is replaced by a single site.
Electrons can tunnel between the sites, and importantly, we include
interactions by means of a Hubbard $U$-term:
\begin{equation}
H = - {\tilde t} \sum_{\sigma} \left( c_{1 \sigma}^{\dagger}
c_{2 \sigma}^{ } + \mbox{h.c.} \right) +
\sum_{i=1}^{2} \left(U n_{i \uparrow} n_{i \downarrow} + E_i(t) n_i \right).
\label{hubbard}
\end{equation}
Here $\tilde t$ is the hopping parameter. In this analysis we
set $\hbar = 1$ and measure all energies in units
of $\tilde t$. $E_i(t)$ is the external electric
potential applied to site $i$. Clearly only the potential
difference, $E_1 - E_2$, is of importance, so we may choose the
convenient parameterisation:
\begin{equation}
E_1(t) = \frac{E}{2} \cos \omega t, \qquad
E_2(t) = -\frac{E}{2} \cos \omega t.
\label{field}
\end{equation}
A numerical investigation of such a model was made recently
in Ref.\cite{zhang} for the case of very weak fields.
The Hilbert space of Hamiltonian (\ref{hubbard}) is six-dimensional,
comprising three singlet states and a triplet. As with the double-dot system,
the singlet and triplet sub-spaces are completely decoupled, allowing
us to study just the singlet space. In the absence of the AC field
the eigenvalues of the singlet Hamiltonian can be found analytically, and
for large $U$ they consist of two almost degenerate excited states, separated
from the ground state by the Hubbard gap $U$. This mimics the eigenvalue
structure of the lowest multiplet of states of the full double-dot system.
We time-evolve the system following the same procedure as before, using
the ground state as the initial state.
As the singlet Hamiltonian is only three-dimensional, however,
the computations can be done correspondingly more rapidly.
In Fig.\ref{interact}b we show the results for \pmin obtained by
setting $U=8$, which clearly shows that this simplified model strikingly
reproduces the behaviour of the full system of Eq.\ref{hamiltonian}.

As the driving field (\ref{field}) is a periodic function of time,
we can make use of Floquet analysis to describe the time evolution
of the system in terms of its Floquet states and quasi-energies \cite{floq}.
The Hamiltonian (\ref{hubbard}) is invariant
under the combined parity operation $x \rightarrow -x; \
t \rightarrow t + T/2$, and so the Floquet states can also
be classified into these parity classes. Close approaches of the Floquet
quasi-energies as the system parameters are varied
produce large modifications of the tunneling rate, and hence of the
system's dynamics. Quasi-energies of different parity classes may cross,
but if they are of the same class they form an anti-crossing.
Numerically the quasi-energies can be conveniently
obtained by diagonalising the time evolution operator for one period
of the driving field, $U(t+T,t)$. This is particularly suited
to the numerical approach we have used, as $U(t+T,t)$ is simply
the operator obtained by time-evolving the identity matrix $I_3$
over one period of the driving field.

We present in Fig.\ref{quasi}a the Floquet quasi-energies as a function
of the field strength for $\omega = 2$, one of the resonant frequencies
visible in Fig.\ref{interact}b. Below the spectrum we also plot the
behaviour of \pmin. We see that the system possesses
{\em two} distinct regimes of behaviour.
For weak fields, as studied in Ref.\cite{zhang},
the Floquet spectrum consists of one isolated
state (which evolves from the ground state) and two states which
make a set of exact crossings. In this regime \pmin decays slowly
to zero, showing little structure. As the field strength exceeds $U$,
however, this abruptly changes to a novel, previously unseen, behaviour
in which \pmin remains close to zero except at a series of narrow peaks,
corresponding to the close approaches of two of the quasi-energies.
A detailed examination of these approaches (see Fig.\ref{quasi}b)
reveals them to be {\em anti-crossings} between the Floquet states
which evolve from the ground state and the higher excited state, and
have the same parity. The remaining state, of opposite parity, makes small
oscillations around zero, but its exact crossings with the other two states
do not correlate with any structure in \pmin.

To interpret this behaviour we seek analytic expressions for
the quasi-energies. We choose to use a perturbational approach
\cite{holthaus}, starting from the Floquet equation:
\begin{equation}
\left( H - i \frac{\partial}{\partial t}  \right)
| \phi_j(t) \rangle = \epsilon_j | \phi_j(t)  \rangle.
\label{station}
\end{equation}
Here $H$ is the full Hamiltonian (\ref{hubbard}),
$\epsilon_j$ are the Floquet quasi-energies, and $| \phi_j(t) \rangle$
are the Floquet states. Our procedure is to first find the
eigenstates of the operator $(H_I - i \frac{\partial}{\partial t})$,
where $H_I$ is the second term in Eq.\ref{hubbard} containing
{\em all} the interaction terms, and then treat the tunneling
component $H_t$ as a perturbation.
An important advantage of this approach is that the Floquet states
are stationary states of (\ref{station}). Consequently, by working
in an extended Hilbert space of $T$-periodic functions \cite{sambe},
the corrections can be evaluated easily by standard Rayleigh-Schr\"odinger
perturbation theory, without requiring more complicated time-dependent methods.

In a real-space representation the interaction terms are diagonal,
and so it can be readily shown that an orthonormal
set of eigenvectors of $(H_I - i \frac{\partial}{\partial t})$ is given by:
\begin{eqnarray}
|&\epsilon&_0(t)\rangle = (\exp \left[i \epsilon_0 t \right], \ 0, \ 0)
\nonumber \\
|&\epsilon&_+(t)\rangle = (0, \ \exp \left[-i (U - \epsilon_+) t
+ i \frac{E}{\omega} \sin \omega t \right], \ 0)
\nonumber \\
|&\epsilon&_-(t)\rangle = (0, \ 0, \ \exp \left[-i (U - \epsilon_-) t
- i \frac{E}{\omega} \sin \omega t \right])
\end{eqnarray}
Imposing $T$-periodic boundary conditions reveals the corresponding 
eigenvalues (modulo $\omega$) to be $\epsilon_0 = 0$ and $\epsilon_{\pm} = U$.
These eigenvalues represent the zeroth-order approximation to the Floquet
quasi-energies, and for frequencies such that $U = n \ \omega$
all three eigenvalues are degenerate.
This degeneracy is lifted by the perturbation $H_t$, and
to first-order the quasi-energies are obtained by diagonalising
the perturbing operator
$P_{ij} = \langle \langle \epsilon_i | H_t | \epsilon_j \rangle \rangle$,
where $\langle \langle \dots \rangle \rangle$ denotes the inner product
in the extended Hilbert space \cite{holthaus}. By using the identity:
\begin{equation}
\exp\left[-i \beta \sin \omega t \right] = \sum_{m=-\infty}^{\infty}
J_m (\beta) \exp \left[-i m \omega t \right],
\end{equation}
to rewrite the form of $|\epsilon_{\pm}(t)\rangle$, the matrix elements of 
$P$ can be obtained straightforwardly, and its eigenvalues subsequently 
found to be $\epsilon_0 = 0$ and $\epsilon_{\pm} = \pm 2 J_n(E/\omega)$.
This solution clearly reduces to the well-known solution for
non-interacting particles when $U = 0$.
Fig.\ref{quasi}a demonstrates the excellent agreement
between this result (with $n = 4$) and the exact quasi-energies
for strong and moderate fields, which allows the position of the peaks
in \pmin to be found by locating the roots of $J_n$. Similar excellent 
agreement occurs at the other resonances.
For weak fields, however, the interaction terms do not
dominate the tunneling terms and the perturbation theory breaks down,
corresponding to the weak-field regime in which \pmin decays smoothly to 
zero. Away from the resonances the first-order correction identically
vanishes, resulting in the lack of structure seen between the
resonant bands in Fig.\ref{interact}b, and it is necessary to go to higher
orders in perturbation theory to obtain the quasi-energy behaviour.

In summary, we have investigated the dynamics of an interacting
two-electron system driven by an AC field. We find that despite the
Coulomb interaction a suitable AC field can nonetheless produce localised
states. This localisation occurs over a range of field strengths at
frequencies for which an integer number of quanta, $n$, 
is equal to the interaction energy. At high fields we find
a novel regime of behaviour in which this localisation vanishes at
the roots of $J_n(E/\omega)$, and
explain this using perturbation theory. These results are of general
applicability to AC-driven systems of interacting electrons, and hold out
the  exciting prospect of controlling and manipulating correlated quantum
states on picosecond timescales by means of applying AC fields.

CEC thanks Sigmund Kohler for numerous stimulating discussions. This
research was supported by the EU via contract FMRX-CT98-0180,
and by the DGES (Spain) through grant PB96-0875.

\bigskip

\bigskip

\begin{references}
\bibitem{kastner} {M.A. Kastner, Phys. Today {\bf 46}, 24 (1993);
R.C. Ashoori, Nature {\bf 379}, 413 (1996).}
\bibitem{shirley} {J.H. Shirley, Phys. Rev. {\bf 138}, B979 (1965).}
\bibitem{floq} {Space does not permit us to give a detailed explanation
of the Floquet approach. Instead we refer the reader to the recent
review article, M. Grifoni and P. H\"anggi, Phys. Rep. {\bf 304},
219 (1998), for an in-depth exposition, and references within.}
\bibitem{hanggi_prl} {F. Grossmann, T. Dittrich, P.Jung and P. H\"anggi,
Phys. Rev. Lett. {\bf 67} 516 (1991).}
\bibitem{holthaus} {M. Holthaus, Z. Phys. B {\bf 59}, 251 (1992).}
\bibitem{hanggi_epl} {F. Grossmann and P. H\"anggi,
Europhys. Lett. {\bf 18}, 571 (1992).}
\bibitem{tamb_epl} {P.I. Tamborenea and H. Metiu, Europhys. Lett. {\bf 53},
776, (2001).}
\bibitem{blick} {R.H. Blick {\it et al.}, Phys. Rev. Lett. {\bf 80}, 4032 (1998).}
%%\bibitem{blick} {R.H. Blick, D. Pfannkuche, R.J. Haug, K. von Klitzing
%%and K. Eberl, Phys. Rev. Lett. {\bf 80}, 4032 (1998).}
\bibitem{ooster} {T.H. Oosterkamp {\it et al.}, Nature (London) {\bf 395}, 873 (1998).}
%%\bibitem{ooster} {T.H. Oosterkamp, T.Fujisawa, W.G. van der Wiel, K. Ishibashi,
%%R.V. Hijman, S. Tarucha and L.P. Kouwenhoven, Nature (London) {\bf 395},
%%873 (1998).}
\bibitem{charles} {C.E. Creffield, W. H\"ausler, J.H. Jefferson and
S. Sarkar, Phys. Rev. B {\bf 59}, 10719 (1999).}
\bibitem{wang} {H. Wang and X.-G. Zhao, J. Phys. Cond. Matt. {\bf 7},
L89 (1995).}
\bibitem{zhang} {P. Zhang and X.-G. Zhao, Phys. Lett. A {\bf 271}, 419
(2000).}
\bibitem{sambe} {H. Sambe, Phys. Rev. A {\bf 7}, 2203 (1973).}
\end{references}
\end{document}